\documentclass[12pt]{article}
\usepackage{amsmath}
\usepackage{graphicx,psfrag,epsf}
\usepackage{enumerate}
\usepackage{natbib}
\usepackage{url} % not crucial - just used below for the URL 

\newtheorem{theorem}{Theorem}

\usepackage{lineno,xcolor}
\begin{document}

\def\spacingset#1{\renewcommand{\baselinestretch}%
{#1}\small\normalsize} \spacingset{1}

%%%%%%%%%%%%%%%%%%%%%%%%%%%%%%%%%%%%%%%%%%%%%%%%%%%%%%%%%%%%%%%%%%%%%%%%%%%%%%

  \title{\bf The Covering Principle: A New Approach to Address Multiplicity in Hypotheses Testing}
  \author{Huajiang Li \\Email: HLi@avanir.com\\
    Avanir Pharmaceuticals, Inc.\\ 30 Enterprise, Aliso Viejo, California, 92656, U.S.A.\\
    and \\
   Hong Zhou\\ Email: hzhou@astate.edu \\
    Department of Mathematics and Statistics, Arkansas State University\\ P.O.Box 70, State University, Arkansas, 72467, U.S.A.}
  \maketitle

\bigskip
\begin{abstract}
The closure and the partitioning principles have been used to build various multiple testing procedures in the past three decades. The essence of these two principles is based on parameter space partitioning.  In this article, we propose a novel approach coined the covering principle from the perspective of rejection region coverage in the sample space. The covering principle divides the whole family of null hypotheses into a few overlapped sub-families when there is a priority of making decisions for hypothesis testing. We have proven that the multiple testing procedure constructed by the covering principle strongly controls the familywise error rate as long as the multiple tests for each sub-familiy strongly control the type I error. We have illustrated the covering principle can be applied to solve the general gate-keeping problems.   
\end{abstract}

\noindent%
\vfill

\newpage
%\spacingset{1.85} % DON'T change the spacing!
\section{Introduction}
%\label{sec:intro}

The key issue in the multiple hypotheses testing is to control the familywise error rate. Commonly, there are a few ways to deal with the multiplicity issue. First is to cut the spending of the significance level $\alpha$ as it does in Bonferroni procedure and its modifications (\citeauthor{Holm79}, \citeyear{Holm79}; \citeauthor{Hoch88}, \citeyear{Hoch88}; \citeauthor{Homm88}, \citeyear{Homm88}; \citeauthor{LMZ17}, \citeyear{LMZ17}).  Second is to plan the order of testing the null hypotheses as it does in the gate-keeping procedures (\citeauthor{DT07}, \citeyear{DT07}; \citeauthor{DTLW08a}, \citeyear{DTLW08a}; \citeauthor{DTW08b}, \citeyear{DTW08b}; \citeauthor{DTB10}, \citeyear{DTB10}; \citeauthor{DT13},\citeyear{DT13}). Third is to make additional assumptions such as the independence among the null hypotheses or the test statistics following the multivariate normal distribution (\citeauthor{TLD98}, \citeyear{TLD98}).

Two important principles: the closure principle \citep{Marcus76} and the partitioning principle (\citeauthor{FS02},\citeyear{FS02}; \citeauthor{Sonn08}, \citeyear{Sonn08}), are widely used to construct various multiple test procedures that can strongly control the familywise error rate.  In the clinical trials, however, the multiple study objectives usually exhibit a hierarchical structure. That is, the objectives can be divided into different tiers according to their importance, namely, primary, secondary, tertiary, and so on. In this article, we introduce a novel principle termed the covering principle for the construction of the multiple testing procedures. The covering principle analyzes the rejection regions in the sample space based on the priorities of the decisions for testing the null hypotheses and divides the whole family of null hypotheses into a few overlapped sub-families, for which either the closure or the partitioning principle can be used. Section \ref{S:cp} introduces the theorem of the covering principle mathematically. Then we will apply the covering principle to a real clinical trial as well as a general gate-keeping problem in Section \ref{S:gate}. The significance and importance using the covering principle are discussed in Section \ref{Disc}. Finally, we will prove Theorem \ref{Th1} that the familywise error rate is strongly controlled for the whole family as long as it is controlled in each sub-family in Appendix.

\section{The Covering Principle}\label{S:cp}

Denote $N=\{1,2,\dots,n\}$ as the index set of a family of $n$ null hypotheses $\{H_1, H_2,\dots,H_n\}$ with the corresponding test functions $\phi=\{\phi_1, \phi_2, \dots, \phi_n\}$ and rejection regions $R_1,R_2,\dots,R_n$. Each $\phi_i$ ($i = 1, 2, \dots, n$) is an elementary test function, where
\[\displaystyle \phi_i=\left\{ \begin{array}{ll}
				1, & \mbox{if $H_i$ is rejected}\\
				0, &\mbox{if $H_i$ is accepted.}
			\end{array}
\right.  \]

For $\emptyset \neq S \subseteq N$, let $\Phi_{\alpha}(S)$ denote the set of all $\alpha$-level multiple tests for the family of null hypotheses with an index set $S$, where $0<\alpha<1$. If $\phi=\{\phi_i: i \in S\} \in \Phi_{\alpha}(S)$, then it indicates the multiple test $\phi$ strongly controls the familywise error rate on $S$ at the significance level $\alpha$. For a group of elementary test functions $\phi_i, i\in S$, define
\[ \displaystyle \min_{\substack{i \in S}} \phi_i = \left\{ 
			\begin{array}{ll}
				1, & \mbox{if $\forall i\in S, \phi_i=1$}\\
				0, & \mbox{otherwise.}
			\end{array}
\right. \]
and
\[\displaystyle  \max_{\substack{i \in S}} \phi_i = \left\{ 
			\begin{array}{ll}
				1, & \mbox{if $\exists i \in S, \phi_i=1$}\\
				0, & \mbox{otherwise.}
			\end{array}
\right.  \]
For any two elementary test functions $\phi_1$ and $\phi_2$, denote $\phi_1 \leq \phi_2$ if $\{\phi_1=1\}$ implies $\{\phi_2=1\}$; in terms of the rejection regions, $R_1\subseteq R_2$.

\begin{theorem}[Covering principle]\label{Th1}
Suppose $\emptyset \neq I \subset N$, $\emptyset \neq J \subset N$, $I \cap J = \emptyset$, and $\bigcup_{i\in I} R_i\subseteq \bigcup_{j\in J} R_j$. Denote
\begin{align}
\phi^j = \{\phi_i^j:i \in N\setminus j \}, &\forall j\in J. \label{phi-j} \\
\phi^{I}=\{\phi_{i}^{I}: i\in N\setminus I \}. & \label{phi-I} 
\end{align}
where $\phi^j$ denote a group of $|J|$ multiple tests which does not include the $j$th hypothesis, $j\in J$. Each $\phi^j$ consists of $|N\setminus j|$ elementary test functions $\phi_i^j, i \in N\setminus j $. Similarly, $\phi^{I}$ denote a multiple test which consists of $|N\setminus I|$ elementary test functions $\phi_{i}^{I}, i\in N\setminus I$, whose indices are not in $I$. Define   
\begin{equation}\label{psi}
\psi_i = \begin{cases}
		\displaystyle	\min(\min_{\substack{j \in J}}\phi_i^j, \phi_i^I), & \text{if $i \in N\setminus I$}   \\ 
		\displaystyle	\min(\min_{\substack{j \in J}}\phi_i^j, \max_{\substack{j \in J}}\psi_j), & \text{if $i \in I$.} 
			\end{cases} 
\end{equation}
If $\phi^j \in \Phi_{\alpha}(N\setminus j)$, $\forall j\in J$ and $\phi^{I} \in \Phi_{\alpha}(N\setminus I)$, then $\{\psi_i:i\in N\}\in \Phi_{\alpha}(N).$
\end{theorem}

The following is an explanation of Theorem \ref{Th1}. Suppose that there exist two nonempty index sets $I\subset N$ and $J\subset N$, $I\cap J=\emptyset$, i.e. these two sets of hypotheses $H_j, j\in J$, and $H_i, i\in I$, are not overlapped. Furthermore, there are orders when these hypotheses are tested. In order for $H_i, i\in I$, to be tested, at least one of the hypothesis $H_j, j\in J$, must be tested and rejected first. For example, the hypotheses $H_j, j\in J$, could be related to the primary endpoints and $H_i, i\in I$, could be related to the secondary endpoints in clinical trials. For parallel gate-keeping, if a  null hypothesis on a secondary endpoint is rejected, then at least one hypothesis on one of primary endpoints already has been rejected. If this is the case, we say the set of hypotheses $H_j, j\in J$, dominates the set of hypotheses $H_i, i\in I$. Simply say, there is a dominance relationship between hypotheses $H_i, i\in I$, and $H_j, j\in J$. From the perspective of hypotheses testing, the orders of testing multiple hypotheses can be illustrated in terms of the logical relationship among their rejection regions: $\bigcup_{i\in I} R_i\subseteq \bigcup_{j\in J} R_j$. 

It may seem that the definition of $\psi_i $ in Equation \ref{psi} is circular. In fact, domains for the index $i$ are mutually exclusive. For the first part of the definition of the test function $\psi_i $, hypotheses $H_i, i\in N\setminus I$, could be either hypotheses $H_j, j\in J$, which dominate $H_i, i\in I$, or independent hypotheses. An independent hypothesis is the one that has no dominance relationship with other hypotheses. The first part of $\psi_i$ defines a test function to reject any hypothesis $H_i$ whose index is not in $I$. That is, $H_i, i\in N\setminus I$, will be rejected if it is rejected in all subsets which contain it. 

The second part of $\psi_i$ in Equation \ref{psi} defines test functions for those hypotheses whose indices are within $I$. These hypotheses $H_i, i\in I$, are dominated by $H_j, j\in J$. In order for any hypothesis $H_i, i\in I$, to be rejected, not only at least one $H_j, j\in J$, must be rejected first, but also $H_i, i\in I$, must be rejected in all subsets which contain it.    

Then, the covering principle in Theorem \ref{Th1} states that the original whole family of $n$ null hypotheses can be decomposed into $|J| +1$ sub-families with index sets $N\setminus I$ and $N\setminus j, \forall j\in J$. In other words, the original multiple testing problem on the family of $n$ null hypotheses with the index set $N$ can be divided into $|J| +1$ multiple testing problems. The corresponding multiple tests are $\phi^I$ with the index set $N\setminus I$ and $\phi^j$ with index sets $N\setminus j, \forall j\in J$. Each subset has fewer null hypotheses than the original family and can be tested using any multiple testing procedure. The multiple testing procedure built on this divide-and-conquer strategy strongly controls the familywise error rate for the whole family at the significance level $\alpha$ if the multiple tests $\phi^I$ and $\phi^j$  can control their familywise error rate at the significance level $\alpha$ for their corresponding subsets. Finally, the decision rule for each individual hypothesis can be reached by summarizing the results as follows:

\emph{Step 1}. $H_i$, $i\in N\setminus I$, will be rejected if $H_i$ is rejected in all decomposed subsets in which $H_i$ is contained;

\emph{Step 2}. $H_i$, $i\in I$, will be rejected if at least one of the null hypotheses $H_j, j\in J$, is rejected first. In addition, $H_i$ must be also rejected in all subsets in which $H_i$ is contained.

Simply speaking, in order for any individual hypothesis $H_i, i\in N$, to be rejected, not only one of its precedent and dominant hypotheses in the hierarchy of the hypotheses must be rejected first, but it must also be rejected in all subsets which contain $H_i$. We will illustrate how to use the covering principle to build a multiple testing procedure in Section \ref{S:gate}.  

The covering principle extends the closure principle to a family of hypotheses with the priority of importance when making decisions. It performs a sample space analysis using the union of rejection regions in contrast to the closure principle using the intersection of hypotheses in the parameter space. The covering principle divides the original family of hypotheses into several sub-families based on the constraints among the relationship of the rejection regions, which is constructed by the usage of the hierarchical structure of the decisions on testing the null hypotheses. The multiple testing problem on the whole family then can be carried out on each sub-family with fewer hypotheses. Furthermore, it can strongly control the familywise error rate for the whole family at the significance level $\alpha$ if the multiple tests for every sub-family control type I error at the significance level $\alpha$.

\section{Applications of covering principle}\label{S:gate}

Dmitrienko et al. (\citeyear{DT07}, \citeyear{DTLW08a},  \citeyear{DTW08b}, \citeyear{DTB10}, \citeyear{DT13}) recently proposed an array of the gate-keeping  procedures including tree-structured, multistage and mixture procedures for the general gate-keeping problems. 

The gate-keeping procedures use the hierarchical structure among the the multiple study objectives. It divides the whole family of the null hypotheses into several ordered sub-families $F_1, \dots,F_n$. $F_i$ serves as the gatekeeper of $F_{i+1}$, that means in order for the null hypotheses in $F_{i+1}$ to be tested $F_i$ must be rejected, i.e. either all the null hypotheses in $F_i$ must be rejected (serial gate-keeping) or at least one null hypothesis in $F_i$ must be rejected (parallel gate-keeping). In other words, if $F_i$ is not rejected, then $F_{i+1}$ will be accepted automatically.

The covering principle provides a very general approach to the gate-keeping problems. The following examples will illustrate how to apply the covering principle to a real clinical trial study using parallel gate-keeping and a more general gate-keeping problem.

\textbf{Example 1:} Cummings et al. (\citeyear{Cumm99}) and Ettinger et al. (\citeyear{Ettinger99}) studied the breast cancer with two primary endpoints: the incidence of vertebral fractures $H_1$ and the incidence of breast cancer $H_2$, and one secondary endpoint: the incidence of non-vertebral fractures $H_3$. The primary endpoint will result in an independent regulatory claim if one of two primary endpoints is effective. That is, the two primary endpoints are parallel, and anyone can serve as the gatekeeper. The test in the secondary endpoint can be carried out as long as at least one of the null hypotheses in the primary family $\{H_1,H_2\}$ is rejected.

\begin{figure}[ht]
  \centering
 % \captionsetup{width=.8\linewidth}
\includegraphics[scale=0.5]{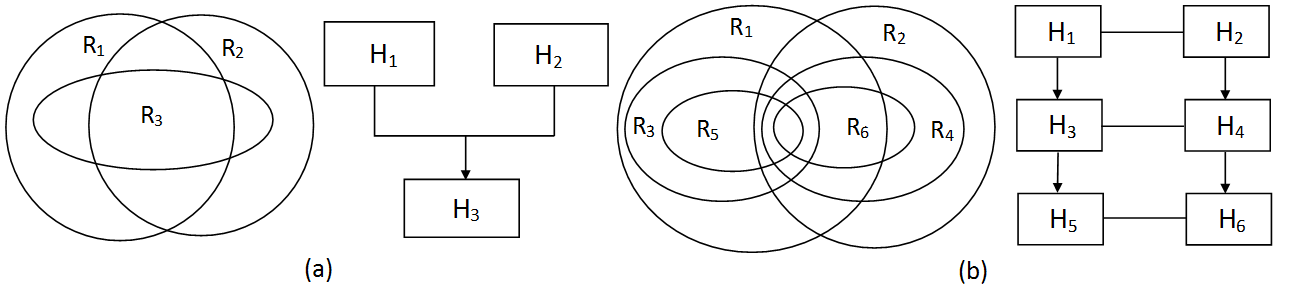}
\caption{(a) Parallel Gate-Keeping (b) General Gate-Keeping}
\label {F1}
\end{figure}

Let $R_1, R_2$ and $R_3$ denote three rejection regions according to null hypotheses $H_1, H_2$ and $H_3$, respectively. Because of the logic constraints among the decisions on testing three null hypotheses, the corresponding rejection regions exhibit the coverage relationship: $R_3\subseteq R_1\cup R_2$ as  illustrated in Figure \ref{F1}(a). According to the covering principle, $N=\{1,2,3\}$, $I=\{3\}$ and $J=\{1,2\}$, the three null hypotheses $H_1, H_2$ and $H_3$ can be divided into $|J|+1=3$ sub-families as follows: $N\setminus I=\{1, 2\}$, $N\setminus 1=\{2, 3\}$, and $N\setminus 2=\{1, 3\}$. The decision rule for each individual hypothesis is as follows:

\emph{Step 1}. $H_1$ will be rejected if it is rejected in sub-families $\{H_1,H_2\}$ and $\{H_1,H_3\}$. Similarly, $H_2$ will be rejected if it is rejected in sub-families $\{H_1,H_2\}$ and $\{H_2,H_3\}$.

\emph{Step 2}. In order to reject $H_3$,  either $H_1$ or $H_2$ must be rejected at \emph{Step 1}, and $H_3$ must also be rejected in sub-families $\{H_2,H_3\}$ and $\{H_1,H_3\}$.

\textbf{Example 2:} Figure \ref{F1}(b) displays a more complicated scenario for which a general gate-keeping procedure can be used. Suppose that three pairs $\{H_1,H_2\}$, $\{H_3, H_4\}$, and $\{H_5,H_6\}$ are parallel to each other and form three tiers. In addition, $\{H_1, H_3, H_5\}$ forms a serial relation as well as $\{H_2, H_4, H_6\}$. In a clinical trial, $H_1$ and $H_2$ may represent hypotheses on the primary endpoints, $H_3$ and $H_4$ on the secondary endpoints, $H_5$ and $H_6$ on the tertiary endpoints. $H_1$, $H_3$, $H_5$ may represent hypotheses on a treatment, $H_2$, $H_4$, $H_6$ on another treatment. Based on the relationship among the decisions on testing the six null hypotheses, the rejection regions exhibit the following coverage relationship: $R_5\subseteq R_3 \subseteq R_1$ and $R_6 \subseteq R_4 \subseteq R_2$.

The covering principle will be used recursively in this example. First, the whole family with six null hypotheses  $\{H_1,H_2,H_3,H_4,H_5,H_6\}$ will be decomposed into $\{H_1,H_2\}$, $\{H_2,H_3,H_4,H_5,H_6\}$, and $\{H_1,H_3,H_4,H_5,H_6\}$ by using the relationship among rejection regions $R_3 \cup R_4\cup R_5 \cup R_6\subseteq R_1 \cup R_2$, that is, $I=\{3,4,5,6\}$ and $J=\{1,2\}$. Second, the sub-family $\{H_2,H_3,H_4,H_5,H_6\}$ will be decomposed into $\{H_2,H_3,H_5\}$ and $\{H_3,H_4,H_5,H_6\}$ by using the relationship $R_6 \cup R_4 \subseteq R_2$, i.e. $I=\{4,6\}$ and $J=\{2\}$. Then,  $\{H_2,H_3,H_5\}$ is decomposed into $\{H_2,H_3\}$ and $\{H_2,H_5\}$ by using $R_5 \subseteq R_3$. Third, $\{H_3,H_4,H_5,H_6\}$ is decomposed into $\{H_3,H_4,H_6\}$ and $\{H_4,H_5,H_6\}$ by using $R_5 \subseteq R_3$;  Fourth, sub-family $\{H_3,H_4,H_6\}$ is further separated into $\{H_3,H_4\}$ and $\{H_3,H_6\}$ by $R_6 \subseteq R_4$. $\{H_4,H_5,H_6\}$ is also separated into $\{H_4,H_5\}$ and $\{H_5,H_6\}$.

Similarly, the sub-family $\{H_1,H_3,H_4,H_5,H_6\}$ can be decomposed into six sub-families: $\{H_1,H_4\}$ , $\{H_1,H_6\}$, $\{H_3,H_4\}$, $\{H_4,H_5\}$, $\{H_3,H_6\}$, $\{H_5,H_6\}$.

Combining all sub-families together, the family of six null hypotheses $\{H_1,H_2,H_3,H_4,H_5,H_6\}$ has been divided into nine sub-families with only two hypotheses each: $\{H_1,H_2\}$, $\{H_1,H_4\}$ , $\{H_1,H_6\}$, $\{H_2,H_3\}$, $\{H_2,H_5\}$, $\{H_3,H_4\}$, $\{H_3,H_6\}$, $\{H_4,H_5\}$,  and $\{H_5,H_6\}$. The original multiple testing problem on the 6-dimension is now reduced into several 2-dimension problems. The decision rule for each hypothesis is as follows:

\emph{Step 1}. Reject $H_1$ if it is rejected in all sub-families: $\{H_1,H_2\}$, $\{H_1,H_4\}$, and $\{H_1,H_6\}$. The rejection of $H_2$ is similar to $H_1$ due to the symmetry.

\emph{Step 2}. Reject $H_3$ if $H_1$ on its the upper level has been rejected, and $H_3$ is also rejected in all sub-families: $\{H_3,H_4\}$, $\{H_3,H_6\}$, and $\{H_2,H_3\}$. The rejection of $H_4$ is similar to $H_3$ due to the symmetry.

\emph{Step 3}. Reject $H_5$ if both $H_1$ and $H_3$ on its upper levels have been rejected, and $H_5$ is also rejected in all sub-families: $\{H_5,H_6\}$, $\{H_2,H_5\}$, and $\{H_4,H_5\}$. The rejection of $H_6$ is similar to $H_5$ due to the symmetry.

\section{Closing Remarks}\label{Disc}

The covering principle can play a key role for the multiple testing problems when there are priorities among the decisions on testing the null hypotheses. It can also be viewed as an extension of two famous closure principle and partitioning principle. Based on the analysis of the rejection regions it decomposed the original family of null hypotheses into a group of sub-families. The merit of this decomposition is threefold.  First, it reduces the dimension of the multiple testing problems since each sub-family has fewer hypotheses, which makes the multiple testing problems much easier at the lower dimensions.  Second, the hierarchical structure among the decisions on testing the null hypotheses as described in the gate-keeping problems are removed by the decomposition. It makes the consolidation of testing results easy and straightforward. Third, any available multiple testing procedure can be used for each sub-family and will be more powerful with fewer hypotheses. After all, the covering principle is very intuitive and easy to understand and use in practice with the help of the flow chart of the decisions and the diagram of the rejection regions.

\section*{Acknowledgement}
The research of Hong Zhou is partially supported by the Arkansas Science and Technology Authority Fund, NO. 15-B-09.

\section{Appendix}\label{Appx}

\subsection*{Proof of Theorem 1}

%\begin{proof}
By the results (\cite{FS02},\cite{Sonn08}), we have $\{\psi_i:i\in N\}\in \Phi_{\alpha}(N)$ if and only if  $\forall \emptyset \neq S \subseteq N$, $\forall \theta \in \displaystyle \cap_{\substack{i\in S}}H_i$, $P_{\theta}(\displaystyle \max_{\substack{i\in S}}\psi_i = 1)\leq \alpha$. For $\forall \emptyset \neq S \subseteq N$, consider two cases of the relationship between $S$ and $J$.

Case I:  $J\not \subseteq S$. There exists a $j_0 \in J$ such that $ j_0\not \in S$, then $S \subseteq N\setminus j_0$. By the definition of $\psi_i$ in equation (\ref{psi}), $\psi_i \leq \displaystyle \min_{\substack{j\in J}}\phi_i^j$, $\forall i\in N$, we have $\psi_i \leq \phi_i^{j_0}$, $j_0 \neq i\in N$. Therefore, $\displaystyle \max_{\substack{i\in S}}\psi_i \leq \displaystyle \max_{\substack{i\in S}}\phi_i^{j_0}$. By the assumption of Theorem \ref{Th1}, $\{\phi_i^{j_0}: i\in N\setminus j_0\} \in \Phi_{\alpha}(N\setminus j_0)$, and $S\subseteq N\setminus j_0$, hence $\forall \theta \in \displaystyle \cap_{\substack{i\in S}}H_i$, $P_{\theta}(\displaystyle \max_{\substack{i\in S}}\phi_i^{j_0}=1)\leq \alpha$.  Therefore, $P_{\theta}(\displaystyle \max_{\substack{i\in S}}\psi_i=1) \leq P_{\theta}(\displaystyle \max_{\substack{i\in S}}\phi_i^{j_0}=1)\leq \alpha$.

Case II: $J\subseteq S$.

If $S\cap I=\emptyset$, then $S=S\setminus I$, hence $\displaystyle \max_{\substack{i\in S}}\psi_i =\displaystyle \max_{\substack{i\in S\setminus I}}\psi_i$.

If $S\cap I \neq \emptyset$, by the definition of $\psi_i$ in equation (\ref{psi}), $\psi_i\leq \displaystyle \max_{\substack{j\in J}}\psi_j, i \in I$,  and since $\forall i \in S\cap I \subseteq I$, hence $\displaystyle \max_{\substack{i\in S\cap I}}\psi_i \leq \displaystyle \max_{\substack{j\in J}}\psi_j$. Because $J \subseteq S$ and $I \cap J = \emptyset$, then $J \subseteq S\setminus I$. Since $S=(S\cap I)\cup (S\setminus I)$, we have $\displaystyle \max_{\substack{i\in S}}\psi_i= \max(\displaystyle \max_{\substack{i\in S\cap I}}\psi_i, \displaystyle \max_{\substack{i\in S\setminus I}}\psi_i)$ $\leq  \max(\displaystyle \max_{\substack{j\in J}}\psi_j, \displaystyle \max_{\substack{i\in S\setminus I}}\psi_i) = \displaystyle \max_{\substack{i\in S\setminus I}}\psi_i$. But $S\setminus I\subseteq S$, hence $\displaystyle \max_{\substack{i\in S}}\psi_i \geq \displaystyle \max_{\substack{i\in S\setminus I}}\psi_i$. Therefore $\displaystyle \max_{\substack{i\in S}}\psi_i =\displaystyle \max_{\substack{i\in S\setminus I}}\psi_i$. 

Similarly, by the definition of $\psi_i$ in equation (\ref{psi}), $\psi_i\leq \phi_i^I, i\in N\setminus I$,  and since $\forall i\in S\setminus I \subseteq N\setminus I$, hence $\displaystyle \max_{\substack{i\in S\setminus I}}\psi_i \leq \displaystyle \max_{\substack{i\in S\setminus I}}\phi_i^I.$ By the assumption of Theorem \ref{Th1}, $\{\phi_i^I: i\in N\setminus I\}\in \Phi_{\alpha}(N\setminus I)$, and $S\setminus I\subseteq N\setminus I$, then $\forall \theta \in \displaystyle \bigcap_{\substack{i\in S \setminus I}}H_i, P_{\theta}(\displaystyle  \max_{i\in S\setminus I}\phi_i^I = 1)\leq \alpha$. Therefore $\forall \theta \in \displaystyle \bigcap_{\substack{i\in S}}H_i \subseteq \displaystyle \bigcap_{\substack{i\in S\setminus I}}H_i, P_{\theta}(\displaystyle  \max_{i\in S}\psi_i = 1) = P_{\theta}(\displaystyle  \max_{i\in S\setminus I}\psi_i = 1) \leq P_{\theta}(\displaystyle  \max_{i\in S\setminus I}\phi_i^I = 1) \leq \alpha$.

Combining Case I and II above, we have $\forall \emptyset \neq S\subseteq N$, $\forall \theta\in \displaystyle \bigcap_{\substack{i\in S}}H_i $, $P_{\theta}(\displaystyle  \max_{i\in S}\psi_i = 1) \leq \alpha$, therefore $\{\psi_i: i\in N\} \in \Phi_{\alpha}(N).$
%\end{proof}

%\bibliographystyle{agsm}
%\bibliography{CP-arxiv-LiZhou}

%\section*{References}

\end{document}